# ANALYSIS OF AN ATTENUATOR ARTIFACT IN AN EXPERIMENTAL ATTACK BY GUNN–ALLISON–ABBOTT AGAINST THE KIRCHHOFF-LAW–JOHNSON-NOISE (KLJN) SECURE KEY EXCHANGE SYSTEM

LASZLO B. KISH [1], ZOLTAN GINGL [2], ROBERT MINGESZ [2], GERGELY VADAI [2], JANUSZ SMULKO [3], CLAES-GÖRAN GRANQVIST [4]

[1] *Department of Electrical Engineering, Texas A&M University, College Station, TX 77843-3128, USA; laszlo.kish@ece.tamu.edu*

[2] *Department of Technical Informatics, University of Szeged, Árpád tér 2, 6720 Szeged, Hungary*

[3] *Faculty of Electronics, Telecommunications and Informatics, Gdansk University of Technology, G. Narutowicza 11/12, 80-233 Gdansk, Poland*

[4] *Department of Engineering Sciences, The Ångström Laboratory, Uppsala University, P.O. Box 534, SE-75121 Uppsala, Sweden*

A recent paper by Gunn–Allison–Abbott (GAA) [L.J. Gunn *et al.*, *Scientific Reports* 4 (2014) 6461] argued that the Kirchhoff-law–Johnson-noise (KLJN) secure key exchange system could experience a severe information leak. Here we refute their results and demonstrate that GAA's arguments ensue from a serious design flaw in their system. Specifically, an attenuator broke the single Kirchhoff-loop into two coupled loops, which is an incorrect operation since the single loop is essential for the security in the KLJN system, and hence GAA's asserted information leak is trivial. Another consequence is that a fully defended KLJN system would not be able to function due to its built-in current-comparison defense against active (invasive) attacks. In this paper we crack GAA's scheme via an elementary current comparison attack which yields negligible error probability for Eve even without averaging over the correlation time of the noise.

## 1. Introduction

Very recently, Gunn–Allison–Abbott (GAA) published a new type of attack [1]—which has been criticized in earlier papers of ours [2–4]—against the Kirchhoff-law–Johnson-noise (KLJN) secure key distribution system [5–11] and asserted, on the grounds of experiments as well as simulation, that an extraordinarily large information leak could occur with cable losses of 0.1 to 1 dB. In particular, GAA found that at a cable loss of 1 dB, and within a fraction of the correlation time of the noise, Eve can extract the key bit with an error probability of around 0.1, which means that her probability of successfully guessing the key bit is $p \approx 0.9$. If this claim were correct it would imply that Eve could separate Alice's and Bob's noises of very different intensities. Even though the validity of the mathematical claim concerning the unconditional security [5] of KLJN would remain intact, GAA's assertions would—if correct—imply that applications of the KLJN scheme would be limited to intra-instrument or inter-chip security.

At face value GAA's claims may seem compelling, but efforts at the Department of Technical Informatics at the University of Szeged in Hungary to reproduce GAA's experiment led to personal communication between GAA and us. Important unpublished details of GAA's experiments were then disclosed, including the display of a serious design deficiency caused by the break-up the Kirchhoff loop in the KLJN system by an attenuator in order to provide the desired loss. This deficiency in GAA's experiments led to their above-mentioned claims, but these claims are flawed and clearly founded on experimental artifacts. It should be emphasized that a complete (*i.e.*, fully defended) KLJN scheme would not have been able to function at all under GAA's conditions, which is a consequence of KLJN's current comparison alarm [13].

Even though the design flaw in GAA's experiment is obvious once their experimental conditions were revealed, it is instructive to demonstrate its seriousness. In this paper we first briefly summarize a number of theoretical arguments why GAA's attack cannot function, and we then analyze their experimental artifact and expose its information leak.

## 2. Why the GAA attack cannot work

We refer to our earlier work on the physical impossibility of low-frequency waves in short cables [3]—which is not directly related to security—and our detailed analysis of GAA's scheme [2], as well as to general arguments [4] about the impossibility of directional couplers. This prior work leads to the following analysis of the problem at hand:

(a) GAA's attack [1] aims at creating a directional coupler to separate the noise components generated by Alice and propagating toward Bob, and generated by Bob and propagating toward Alice, in the KLJN scheme.

(b) The GAA model assumes reflections and propagation at the electromagnetic phase velocity $c_p$ in the cable, which is justified only for waves. However, this is not a serious problem since linear response theory allows one to divide the low-frequency signal into short spikes for which the wave equation does work, to study the response to these spikes separately (including reflected waves propagating at $c_p$), and then to sum up these responses.

(c) The wave based d'Alembert representation used by GAA [1] is well known and works for short transient signals. However, a directional coupler will provide the required output only as long as mixing of forward-going and reflected signals does not occur.

(d) For low-frequency signals, *i.e.*, in the quasi-stationary limit, reflections create a mixture of signals injected at the two ends of the cable. This leads to an *effective phase velocity* $c_{pe}$, which is proportional to the resistance terminating the end toward which the wave propagates, as demonstrated by computer simulations in our earlier work [3]. It should be noted that $c_{pe}$ equals $c_p$ when the terminating resistor is equal to the wave impedance (which is 50 Ω in GAA's work [1,3]).

(e) However, GAA's method [1] requires knowledge of $c_{pe}$ in both directions, which in its turn necessitates information on Alice's and Bob's resistance values. Consequently Eve cannot separate Alice's and Bob's signals unless she knows their magnitudes [2].

(f) Our earlier work [2] analyzed feature (e) mathematically and showed that if Eve, while attempting to extract Bob's noise, assumed the correct value of $c_{pe}$ for wave propagation toward Bob she could indeed distinguish between Alice's and Bob's signals. However, if Eve assumed the wrong $c_{pe}$ value, *i.e.*, the value for wave propagation toward Alice, then she would extract a non-existent noise that had the *same mean-square value as Alice's noise*. In other words, Eve gets what she assumes: if she assumes a termination with the *high* resistance then her evaluated noise will agree with that assumption, but if she assumes a termination with the *low* resistance then her evaluated noise will agree with that assumption instead. Therefore Eve's one-bit uncertainty persists within the GAA method.

(g) GAA's theoretical analysis and computer simulation results are in accordance with the above facts in (f), because both of them show that Eve *cannot extract any information from a lossless cable* [1]. Furthermore, GAA's computer simulations for the smallest loss of 0.01 dB (see Fig. 3 in their work [1]) were in accordance with earlier experimental tests [10] of the KLJN system, which showed similar losses and thus a minor information leak that, however, could be remedied by simple privacy amplification [12].

(h) The *only* situations for which GAA were able to extract information were the set-ups with *losses*. This fact indicates that propagation effects were not the cause of the measured information leak but some other phenomenon. Unfortunately GAA did not show any cable-length-dependent information leak, which might have been used to test the real role of propagation delay effects. According to the information we received from GAA, there remained a significant loss, corresponding to 0.1 dB cable loss, in their system even at the experimental situation indicated as "zero loss", which is not immediately obvious from their paper [1]. Therefore GAA imply by "zero loss" that no additional attenuator was used to increase the losses.

(i) GAA mention [1] that our recent defense method against wire resistance based attacks [14] protects also against the GAA attack. However, this defense method—originally developed against the Second Law attack—boosts the noise temperature at the lower resistance end [14]. The effectiveness of this defense method against GAA's attack is another indication of the irrelevance of the propagation effects in GAA's results.

In conclusion, both our own [3] and GAA's [1] theoretical analyses agree in that no information leak exists in the loss-free case whereas propagation delays are present. In the next section we will show that GAA's experimental results, with attenuators to produce their required loss, contain a severe artifact which would have prohibited the functioning of a complete KLJN system as a consequence of its current-comparison defense against active attacks [13]. We show that Eve can extract the information by elementary measurements with very low error rate, virtually immediately, even without using GAA's statistical tool [1].

## 3. Attenuator artifact in the GAA experiments, and its analysis

The attenuator is a symmetrized voltage divider which provides not only attenuation but also 50 Ω input impedance when the far end of the cable is terminated by 50 Ω. Figure 1 shows the 1 dB attenuator (with approximate values) and its incorporation into the KLJN loop. It is obvious that the shunt resistor $R_2$ breaks the originally single Kirchhoff loop into two loops with a common side. This is an incorrect realization of the KLJN system, because security has been guaranteed only for a single loop involving Alice and Bob. The violation inherent in GAA's circuit is very significant, because the value of $R_2$ (500 Ω) is 20 times smaller than that of Bob's resistor $R_B$ (10 kΩ) and two times smaller than Alice's resistor $R_A$ (1 kΩ). These data prove that GAA's experiments, in the cases with higher losses and information leak, were not conducted on the KLJN system but on something else.

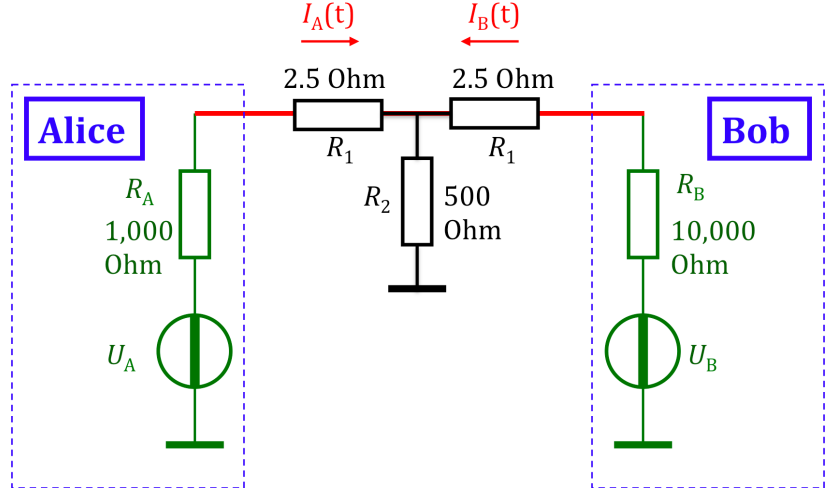

**Figure 1.** Circuit diagram for the 1 dB attenuator applied by GAA to crack the KLJN scheme. *U*, *R*, *I* and *t* denote voltage, resistance, current and time, respectively. GAA's 0.1 dB attenuator had the same structure with appropriate resistances.

We show next that the time-dependent currents $I_A(t)$ and $I_B(t)$ of Alice and Bob, respectively, instead of being equal, as required for security, are strongly unbalanced, and Alice's mean-square current is about five times larger than Bob's. Straightforward circuit noise analysis provides the mean-square currents of Alice and Bob according to

$$\langle I_A^2(t) \rangle = S_{iA}(f) B_{kljn} \simeq \frac{4kT_{eff} R_A}{\left[ R_A + R_B R_2 / (R_B + R_2) \right]^2} B_{kljn}$$

$$+ \left( \frac{R_A^{-1}}{R_A^{-1} + R_2^{-1}} \right)^2 \frac{4kT_{eff} R_B}{\left[ R_B + R_A R_2 / (R_A + R_2) \right]^2} B_{kljn} \qquad (1)$$

and

$$\langle I_B^2(t) \rangle = S_{iB}(f) B_{kljn} \simeq \frac{4kT_{eff} R_B}{\left[ R_B + R_A R_2 / (R_A + R_2) \right]^2} B_{kljn}$$

$$+ \left( \frac{R_B^{-1}}{R_B^{-1} + R_2^{-1}} \right)^2 \frac{4kT_{eff} R_A}{\left[ R_A + R_B R_2 / (R_B + R_2) \right]^2} B_{kljn} \quad , \qquad (2)$$

where it is assumed that the resistors $R_1$ are much smaller than the total resistances of the loops; $S_{iA}(f)$ and $S_{iB}(f)$ are (white) noise spectra for Alice's and Bob's currents; and

*Analysis of a design flaw in the Gunn-Allison-Abbott experiments concerning the KLJN scheme*

$B_{kljn}$ and $T_{eff}$ are noise bandwidth and effective noise temperature of the generators, respectively. Substituting the practical values used in GAA's experiments, we obtain that

$$\frac{\langle I_A^2(t) \rangle}{\langle I_B^2(t) \rangle} = 4.95 \quad , \tag{3}$$

which means that Alice's mean-square current is about five times stronger than Bob's. This extraordinary difference means that even the simplest comparison methods can extract information so that GAA's complex statistical tool [1] is unnecessary. To illustrate this fact, we show below that a simple current comparison, without making statistics or averaging, is sufficient to create an efficient attack.

Eve's task, in order to extract the bit, is to guess which mean-square current is the larger one: Alice's or Bob's. According to Eq. (3) this task is equivalent to guessing, from a few measurement samples, if a current $I_1(t)$ with unit mean-square value

$$\langle I_1^2(t) \rangle = 1 \quad , \tag{4}$$

or another current $I_2(t)$ with mean-square value

$$\langle I_2^2(t) \rangle = 4.95 \quad , \tag{5}$$

is the larger. Without loss of generality, we use this example and assume that the values of the measurement noise currents are already normalized in accordance with Eqs. (4) and (5), which is straightforward because the magnitudes of $\langle I_A^2(t) \rangle$ and $\langle I_B^2(t) \rangle$ are theoretically known by Eve since all the resistors and the effective temperature are public knowledge [4–6].

The simplest protocol is to perform a single measurement on the currents and compare their square with the threshold 4.95. If one of the currents squared is greater than the threshold and the other is smaller, then Eve concludes that the first one is $I_2(t)$.

We obtain the following probabilities about the behavior of the square of a *single measurement value* of the current:

$$\begin{aligned}
P\!\left(I_1^2(t) < 4.95\right) &= F_1(4.95) = 0.974 \\
P\!\left(I_1^2(t) > 4.95\right) &= 1 - F_1(4.95) = 0.026 \\
P\!\left(I_2^2(t) < 4.95\right) &= F_2(4.95) = F_1(1) = 0.68 \\
P\!\left(I_2^2(t) > 4.95\right) &= 1 - F_2(4.95) = 1 - F_1(1) = 0.32
\end{aligned} \tag{6}$$

where $F_1$ and $F_2$ are the chi-squared distributions of $I_1^2(t)$ and $I_2^2(t)$, respectively, with one degree of freedom.

The probability of successful (but not necessarily error-free) guessing, *i.e.*, $I_1^2(t) < 4.95$ and $I_2^2(t) > 4.95$, is

$$P_s = 0.974 * 0.32 = 0.31 \ . \tag{7}$$

In this case, Eve's guess is $\langle I_1^2(t) \rangle < \langle I_2^2(t) \rangle$. The error probability of this guess is

$$P_\varepsilon = P(I_1^2 > 4.95) \, P(I_2^2 < 4.95) = 0.026 * 0.68 = 0.018 \ , \tag{7}$$

which is less than 2%, thus indicating over 98% fidelity of Eve's successful guessing.

It is obvious that the probability of "there is no answer"—*i.e*, when both measured values are below the threshold or when both of them are above the threshold—is given as $0.974*0.68+0.026*0.32 \approx 0.67$. Thus on the average three measurements are needed to get an answer which will have over 98% fidelity. Three independent measurements can be done in three correlation times and yield Eve's 1.8% error probability. Very interestingly, GAA's computer simulations gave the same error probability in three correlation times by using their demanding statistical method as our elementary non-optimized method provides, as apparent from Fig. 3 in their work [1]. (Note that the threshold is at the mean-square value of $I_2$, and much greater than the mean-square value of $I_1$, and therefore a successful measurement is expected to happen with near-unity probability during the correlation time of the noise, because the square of a Gaussian noise typically goes through virtually all values between zero and its mean-square value during the correlation time.)

The *ad hoc* and non-optimized protocol described above, and its small error probability, serve as an illustration of the astronomically large information leak that is caused by the attenuator artifact in GAA's work [1].

## 4. Conclusion

A recent paper by Gunn–Allison–Abbott [1] claimed that the KLJN secure key exchange system—referred to by them as the "Kish key distribution system"—could display a devastating information leak. The present paper refuted their results and showed that GAA's arguments arise from a serious deficiency in the design of their system. Specifically, GAA used an attenuator which broke the single Kirchhoff-loop, which is an essential feature for the security in the KLJN system. Therefore GAA's alleged information leak is trivial. We also cracked GAA's scheme via an elementary current-comparison attack, which yielded negligible error probability for Eve in a short time of the order of the correlation time of the noise.

**Acknowledgements**

We are grateful to Lachlan Gunn for sharing unpublished details on GAA's experimental realization of the KLJN scheme and thereby helping us to identify the design artifact that yielded the major information leak in the attack described in GAA's paper.

LBK's visit of the University of Szeged (October 6-17, 2014) was supported by Texas A&M University. RM's, ZG's and GV's research was partially supported by the European Union and the State of Hungary, co-financed by the European Social Fund in the framework of TÁMOP 4.2.4. A/2-11-1-2012-0001 'National Excellence Program'. JS's research was partially supported by the Statutory Funds of Gdansk University of Technology, Faculty of Electronics Telecommunications and Informatics for JS.